\documentclass[american,aps,pra,reprint,superscriptaddress,twocolumn,showpacs]{revtex4-1} 
\usepackage[unicode=true,pdfusetitle, 
bookmarks=true,bookmarksnumbered=false,bookmarksopen=false, 
breaklinks=false,pdfborder={0 0 0},backref=false,colorlinks] {hyperref}
\hypersetup{ 
colorlinks,linkcolor=myurlcolor,citecolor=myurlcolor,urlcolor=myurlcolor}
\usepackage{braket,colortbl,cleveref,amsthm,amsmath,amssymb,txfonts}
\definecolor{myurlcolor}{rgb}{0.0,0.0,0.5}
\usepackage{graphics,graphicx}
\usepackage{color}
\usepackage{graphicx}
\usepackage[format = plain,labelfont = bf,up, textfont = normal , up, 
justification =raggedright, singlelinecheck =false]{caption}
\usepackage{tabularx}
\usepackage{subfigure}
\usepackage{amsmath}
\DeclareMathOperator{\tr}{Tr}
\usepackage{graphicx}
\usepackage[utf8x]{inputenc}
\usepackage{color}
\usepackage{amsmath}
\usepackage{braket}
\usepackage{latexsym}
\usepackage{amssymb}
\usepackage{amsthm}
\usepackage{bm}
\usepackage{graphics,epstopdf}
\usepackage{color}\usepackage{amsmath}
\usepackage{enumitem}
\usepackage{fmtcount}
\usepackage{booktabs}
 
\usepackage{epsfig}

\theoremstyle{plain}
\def\bea{\begin{eqnarray}}
\def\eea{\end{eqnarray}}
\def\ba{\begin{array}}
\def\ea{\end{array}}

\def\ket{\rangle}
\def\bra{\langle}
\def\beq{\begin{equation}}
\def\eeq{\end{equation}}

 \newtheorem{theorem}{}
  \newtheorem*{theorem*}{}
  \pdfoutput=1
\begin{document}

\title{Quantum speed limit constraints on a  nanoscale autonomous refrigerator}

\author{Chiranjib Mukhopadhyay}
\email{chiranjibmukhopadhyay@hri.res.in}
\affiliation{Harish-Chandra Research Institute, HBNI, Allahabad 211019, India}

 \author{Avijit Misra}
 \affiliation{Optics and Quantum Information Group, The Institute of 
Mathematical Sciences, HBNI, Chennai 600113, 
India}
\author{Samyadeb Bhattacharya}
\affiliation{Harish-Chandra Research Institute, HBNI, Allahabad 211019, India}
\affiliation{S. N. Bose National Centre for Basic Sciences, Kolkata-700106, India}

 \author{Arun Kumar Pati}
\affiliation{Harish-Chandra Research Institute, HBNI, Allahabad 211019, India}

%
%
%

\begin{abstract} 
Quantum speed limit, furnishing a lower bound on the required time for the  evolution of a quantum system through the state space, imposes an ultimate natural limitation to the dynamics of physical devices. Quantum absorption refrigerators, on the other hand, have attracted a great deal of attention in the last few years.  In this paper, we discuss  the effects of quantum speed limit on the performance of a quantum absorption refrigerator. In particular, we show that 
there  exists a trade-off relation between the steady cooling rate of the refrigerator and the minimum time taken to reach the steady state. Based on this, we define a figure of merit called ``bounding second order cooling rate" and show that this scales linearly with 
the unitary interaction strength among the constituent qubits. We also study the increase of bounding second order cooling rate with the thermalization strength. We subsequently demonstrate that coherence in the initial three qubit system can significantly increase the bounding second order cooling rate. We  study the efficiency of the refrigerator at maximum bounding second order cooling rate and,  in a limiting case, we show that the efficiency at maximum bounding second order cooling rate is given by a simple formula resembling the Curzon-Ahlborn relation.


 
\end{abstract}

\maketitle

\section{Introduction}
Since Sadi Carnot proposed an ultimate quantitative bound 
\citep{carnot} on the performance of thermal devices, our understanding of 
microscopic physics has been revolutionized over the last two centuries. 
In contradistinction with the grounding of classical thermodynamics on a deterministic microscopic dynamical model, small scale thermal machines with constituents obeying laws of quantum theory have only recently come within our purview \citep{huberreview, kosloffreview}. However, these quantum machines still run up against the Carnot bound and barring non-thermal reservoirs like coherent \citep{nonthermal1} or squeezed baths \citep{nonthermal2,nonthermal3,nonthermal4,nonthermal5}, generally fail to surpass it.  Yet, the Carnot bound, attainable only through an infinitely slow process, is of limited practical utility compared to the performance of thermal devices at finite power. Classically, a bound on the efficiency at maximum power for cyclic thermal engines\citep{reitlinger,novikov,ca} and refrigerators \citep{yan,yanchen,velasco} was obtained in several works, providing impetus to a growing body of research on finite time and non-equilibrium thermodynamic processes \citep{wubook,prigoginebook}. Investigations on finite time cyclic thermodynamic processes for quantum thermal machines \citep{abah, kurizkireview,raamuzdin} have already provided crucial physical insight on topics  ranging from the origin of friction\citep{friction1,friction2,friction3,friction4} in thermal processes to the third law of thermodynamics \citep{thirdlaw1,thirdlaw2}. 
\\
Apart from cyclic thermal machines; one may also conceive of self-contained thermal machines where the energy required to perform a task, e.g. cooling a cold bath, is provided by a third bath, thus requiring no external control. Motivated by algorithmic cooling, such a  quantum absorption refrigerator (QAR) was proposed in \citep{smallest}.  Numerous other proposals of QAR have already been put forward \citep{abs1,abs5, adesso1, rc_silva}. Experimental realization of QAR has been recently achieved  in ion trap systems \citep{iontrap} . Experimental proposals for various alternate systems, viz. cavity QED systems \citep{abs4} and quantum dots \citep{abs3}, have also been made.  In the case of autonomous three-qubit refrigerators, the constituent qubits interact among themselves
via an energy conserving unitary operation as well as thermalize with their corresponding heat baths.
Depending upon the initial conditions, the setup at its steady state may be shown to act as a refrigerator which cools down the cold bath by extracting heat at a constant rate.  These machines have the advantage of both simplicity as well as practical usefulness. \\ In addition to finding the steady state performance of such refrigerators, one should also consider how long it takes for the system to reach the steady state. If the system cools very reliably at steady state but only gets there very slowly, it is of limited utility.  One can study the finite time characteristics of the QAR \citep{brask15,  sreetama, Mitchison} for cooling the cold qubit, however this approach may necessitate precise time control mechanism for decoupling the cold qubit. In addition, our goal is steady cooling  of the cold bath rather than cooling the qubit by itself.  In this paper, we seek to understand the behaviour of QAR conditioned through the intrinsic restriction to evolution of a quantum system through its state space. The latter feature, known as the quantum speed limit is a fundamental feature of quantum dynamics and finds several applications  in quantum computation, control theoretic settings and in the study of shortcut towards adiabaticity (See,  for example,  Ref. \citep{deffnerreview}. and references therein for an overview). To this end, we analyze a 
figure of merit encompassing this feature of the QAR described in \citep{smallest}. This 
figure of merit, called \emph{``bounding second order cooling rate" (BSOCR)} is defined as the product of the equilibrium cooling rate 
and the maximum possible speed of attaining equilibrium.   This is of significant practical importance, since even if a refrigerator cools the cold bath very well in the steady regime, it will be of limited use if it takes a very long time to reach the steady regime. We link the proposed figure of merit with the transient features of the refrigerator and go on to illustrate the dependence of BSOCR on various system parameters. We show that BSOCR increases linearly with the coupling strength.  We note that the presence of quantumness in the form of initial coherence in the system can boost the value of BSOCR. This is followed up by an investigation of the figure of merit vis-a-vis the refrigeration efficiency. We show that at a high temperature limit and subject to other specific conditions, we can recover an expression for efficiency at maximum value of BSOCR which scales similarly to the Curzon-Ahlborn bound.

The paper is organized as follows. In section II, we briefly recapitulate the QAR model used in our work. In section III, we mention the form of quantum speed limit used. Our novel results are presented in section IV. In section V, we conclude through a discussion of our results and possible future extensions thereof.  The detailed calculations are to be found in the appendix.


%

\section{Absorption refrigerator at steady state}
We consider the model of three qubit absorption refrigerator introduced in 
Ref. \citep{smallest}. In what follows we briefly discuss about the model and its working principle. The three qubits consisting the refrigerator are coupled to three different baths at 
different temperatures. The first qubit which is  the object to be cooled, is coupled to the coldest bath at temperature $T_c$. The second qubit which takes energy from
the first qubit and disposes into the environvent, is coupled to a hotter bath at temperature $T_r$. The third and final qubit which provides the free energy for refrigeration 
is coulped to the hottest bath at temperature $T_h$. Here $T_c\leq T_r\leq T_h.$
 Without loss of generality, the ground state energy of all the qubits are considered to be zero and the excited state energy 
of the $ i$-th  qubit is $ E_{i} $, where $i\in \{c,r,h\}$. The free Hamiltonian of the combined system 
is $ H_{0}=\sum_{i\in \{c,r,h\}} E_{i} |1\rangle_i\langle1|$. 
 In thermal equilibrium the qubits are in the corresponding thermal states
 $\tau_{i}=r_{i}\Ket{0}\Bra{0}+\bar{r}_{i}\Ket{1}\Bra{1} $, where \beq 
r_{i}=(1+e^{-\beta_i E_{i}})^{-1} \label{popul} \eeq is the probability of the $i$-th qubit to 
be in the ground state and $\bar{r}_{i} = 1 - r_{i}$. Here $\beta$ is the inverse temperature $1/T$. {\color{black}The populations are thus dependent on both the temperature and energy spacing.} The Boltzmann costant $k_B$ is set to unity and this convention is followed through out the paper. 
The total system is initially in a product state of locally thermal qubits $\rho_{0}=\tau_{c}\otimes\tau_{r} \otimes \tau_{h}$. 
The qubits interact via the  interacting hamiltonian $H_{int}=g(\Ket{101}\Bra{010}+\Ket{010}\Bra{101})$. Here the order of the qubits is $c,r,h$ which is maintained throughout the paper unless  otherwise mentioned.
The interaction strength $g$ is taken weak enough compared to the 
the energy level $E_i$-s, i.e., $ g<<E_{i} $, so that
the energy levels and the energy eigenstates of the combined system are almost unaltered and the temperature of the each qubit can be defined neglecting the interaction energy \citep{smallest}.
The total Hamiltonian of the combined system is  given by 
\begin{equation}
 H=\sum_{i\in \{c,r,h\}}E_i |1\rangle_i\langle1|+ g(\Ket{101}\Bra{010}+\Ket{010}\Bra{101}).
\end{equation}

As the qubits are coupled with heat baths at each time step there is a finite probability that it will thermalize. Suppose $ p_{i} $ 
is the probability density per unit time that the $ i$-th qubit will thermalize back. 
Then the evolution of the combined system is given by the following master eqaution
\begin{equation}
\label{master}
 \frac{\partial \rho}{\partial t}  = -i[H,\rho]+\sum_{i\in \{c,r,h\}}  
p_{i} (\tau_{i}\otimes \tr_{i}\rho - \rho).
\end{equation}
It is necessary to mention that this master equation is valid only in the perturbative regime where the simultaneous thermalization of more than one qubit can be neglected. The steady state refrigeration with the aforementioned model has been demonstrated in great detail in Ref. \cite{smallest, jpa}. 
The steady state for this master equation, as obtained in Ref. \citep{jpa}, is given 
as 

\begin{equation}
\label{final-state}
\rho_{f} = \tau_{c} \otimes \tau_{r} \otimes \tau_{h} + \gamma \sigma_{crh},
\end{equation} 
where 
\begin{eqnarray}
 \sigma_{crh} &=&\Big(Q_{rh} Z_c\tau_r\tau_h + Q_{ch} \tau_c Z_r \tau_h + Q_{cr} \tau_c \tau_r Z_h\  \nonumber \\
  && + q_c \tau_c Z_{rh} + q_r \tau_r Z_{ch} + q_h Z_{cr}\tau_h + Z_{crh} + \frac{q}{2g}Y_{crh}\Big).
\end{eqnarray}
Here, $Y_{crh}~=~i|101\rangle\langle010|-i|010\rangle\langle101|$ and $Z_{crh} = |010\rangle\langle010|-|101\rangle\langle101|$, $Z_{ij} = \tr_k Z_{ijk}$, where $\{i,j,k\}\in\{c,r,h\}$ and  $q_i$ and $Q_{jk}$, 
are given as follows $ q_i = \frac{p_i}{q-p_i}$, $Q_{jk} = \frac{p_jq_k + p_kq_j}{q-p_j-p_k}$,
where $q = p_c + p_r + p_h$. The parameter $\gamma$ in Eq. \ref{final-state} is given by
\begin{equation}\label{e:gamma}
  \gamma = \frac{-\Delta}{2 + \frac{q^2}{2g^2} + \sum_i q_i + \sum_{jk} Q_{jk}\Omega_{jk}}
\end{equation}
where
$\Delta = r_1 (1-r_2) r_3 - (1-r_1) r_2 (1-r_3), \quad \Omega_{jk} = r_j'(1-r_k') + (1-r_j')r_k.$
Here $r'_i = (1-r_i)$ for $i=r$, otherwise $r'_i = {r_i}$.
 The cooling 
rate for the cold bath is given as \beq {Q_{c} = q\gamma E_{c}, \label{qc}}\eeq which clearly shows that the machine acts as a 
refrigerator only when $\gamma > 0$. It has  been shown that the efficiency $\eta$ of this refrigerator is equal to $E_c / E_h $ \citep{jpa}.

\section{Quantum speed limit}
Quantum evolution for a closed system is a unitary map. It has been shown that the fluctuation \citep{mandelstam,anandan} or the average value
\citep{margolus, lloyd} of the generator of such maps determines the 
maximum rate of unitary evolution of a quantum system through the corresponding state space, giving rise to the concept of a limiting speed for dynamical evolution. For pure quantum states, this speed of evolution was introduced by Anandan and Aharonov \citep{anandan} utilizing the Fubini-Study metric, with subsequent works \citep{pati, vaidman} building on the concept. Generalizations for  mixed states in the case of unitary evolutions were proposed \citep{giovanetti} and tightness of bounds found earlier were proved in some cases \citep{toffoli}.  For a generalized quantum 
evolution characterized by CPTP maps, it is possible to find similar speed 
limits \citep{debamixed, deffnerlutz,Debasis, taddei, diego}. In particular, for a Markovian channel on an open quantum system 
expressed via a dynamical subgroup $\mathcal{L}$, the following lower bound on 
the time $t_{evolution}$ required for evolution of a quantum system from initial 
state $\rho_{0}$ to a state $\rho_{f}$ was given in Ref. 
\citep{delcampo} as

\begin{equation}
\label{delcampo}
t_{evolution} \geq  \frac{|\cos \theta -1| \tr \rho_{0}^{2}}{\sqrt{ \tr  
\left(\mathcal{L}^{\dagger} \rho_{0} \right)^{2}}} = \tau,
\end{equation}

\noindent where $\theta = \cos^{-1} \frac{\tr \left(\rho_{0} \rho_{f} \right)}{ \tr 
\left(\rho_{0}^{2} \right)}$ is expressed in terms of relative purity between the initial and the final state. Thus, $1/\tau$ can be interpreted as the maximum speed of the evolution. In this paper,
using this maximum speed of evolution, we show how the same constrains the performance of the 
QAR.

\section{Effects on the performance of quantum refrigerator from quantum speed limit}
In this section, we establish a link between the cooling rate of the QAR at its steady state and the minimum
time that it takes to reach the steady state. We define the novel figure of merit, i.e.,  the product of  the steady cooling power of the QAR and the maximum speed of evolution to the steady state as	

\begin{equation}
\label{coolingrate1}
\chi = \frac{Q_{c}}{\tau}.
\end{equation}
Note that, for better performance, we need higher 
$\chi$, i.e., higher cooling rate as well as faster evolution to the steady state. Interestingly, we observe
that there lies a trade-off between these two desired criteria.
It will be interesting to explore in detail how the performance of steady QAR depends 
on $\chi$, as well as how $\chi$ itself depends on the system parameters.  Before doing so, we would like to digress a bit towards transient refrigeration by QAR. 

The efficiency and cooling rate in the transient regime by QAR have been extensively studied in Ref. \cite{sreetama}. The figure of merit $\chi$ is the upper bound on the temporal average of time derivatives of the instantaneous cooling rate. The latter 
quantity, say $P(t)$, is defined as $P(t) = \frac{dQ_{c}(t)}{dt}$, where $Q_{c}(t)$ is 
the instantaneous cooling rate. Now, the time average \beq  \overline{P (t)} = 
\frac{1}{t_{evolution}} \int_{0}^{t_{evolution}} \frac{dQ_{c}(t')}{dt'} dt' = 
\frac{Q_{c}}{t_{evolution}} \leq \frac{Q_{c}}{\tau} = \chi,  \eeq justifying our 
assertion. {\color{black} Thus the time-derivative of the transient cooling rate, averaged over the entire duration of dynamics, is upper bounded by our figure of merit. The transient cooling rate is, of course, the amount of heat drawn from the bath per unit time during the transient regime. Thus, the time derivative of cooling rate may be argued as a second order time-derivative of the amount of heat drawn from the bath during transient dynamics. Since our figure of merit represents an upper bound on the average time-derivative of the transient cooling rate, we coin the term \emph{bounding second order cooling rate (BSOCR)} for the figure of merit $\chi$. To avoid misleading the reader, we emphasize that the exact expression for $\chi$ is homogeneous to $Q_c$, i.e., the steady cooling rate, and not to the time derivative of the transient cooling rate. }  

\begin{figure}
\includegraphics[scale=0.25]{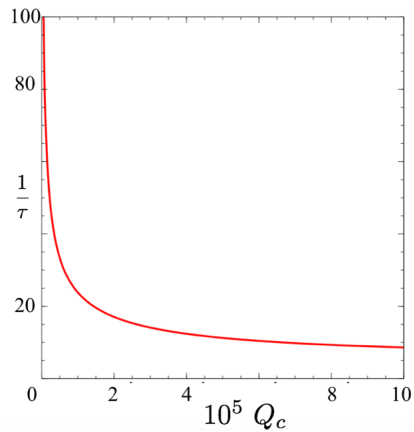}
\caption{ {\color{black}Demonstration that high equilibrium cooling power $Q_c$ can come at the cost of slow evolution to equilibrium state. When all parameters other than $g$ are fixed, steady cooling rate $Q_c$ and  the inverse of evolution time $1/ \tau$, both functions of $g$, are parametrically plotted by varying the interaction strength $g$ from $10^{-4}$ to $10^{-1}$. The parameters are chosen as $E_c = 1$, $p_c = p_r = p_h = 0.1, \eta = 1, T_c =1,T_r = 5,$ $ T_h = 10$.}}
\label{motivation}
\end{figure}

Let us now see how the BSOCR controls the steady refrigeration. In terms of the steady state parameters, the BSOCR can be expressed in terms of the initial state $\rho_0$  via  \eqref{final-state}, \eqref{e:gamma}, and \eqref{delcampo} as
\begin{equation}
\label{trade-off}
 \frac{Q_{c}}{\tau} = \frac{q E_{c}}{ \left[\tr (\rho_{0} 
\sigma_{crh})\right]/ \left[\sqrt{ \tr ([H_{int}, \rho_{0} 
])^{2} } \right]}  
\end{equation}
Using the equation above, we demonstrate the trade-off between the steady cooling power and the maximum speed of evolution. As $Q_c$
and $1/\tau$ both are function of the interaction strength $g$, we plot these quantities by varying the interaction strength $g$. Fig. \ref{motivation}  clearly demonstrates the trade-off between steady cooling power $Q_c$ and maximum speed of evolution
$1/\tau$, implying one can only get better cooling rate by compromising on the speed of evolution. This sets up the main crux of our work. It is also surprising that the performance of an autonomous quantum thermal machine at the steady state depends on the minimum time taken to reach the steady state. In the rest of the work, we shall study how the BSOCR $\chi$ depends on the parameters of the QAR. We shall also investigate whether initial quantumness can enhance the performance of the steady QAR.

\subsection{Interaction strength vs BSOCR}
We know that the three-body interaction cools the cold qubit and draws heat from the cold bath. Hence, the more the
interaction strength, the stronger the biasing 
facilitating the refrigeration. However, the cooling power depends in a complicated way on the strength $g$ of the 
three-body interaction. On the contrary, from Eq. \ref{trade-off} (see the appendix  for a detailed expression) when the resetting probabilities
$\{p_i\}$ are equal),  we observe that \textit{the BSOCR grows linearly with the strength $g$ of the three-body interaction 
$H_{int}$ when the qubits start from their respective thermal states.}
\begin{figure}
\centering
    \includegraphics[width=0.23\textwidth, keepaspectratio]{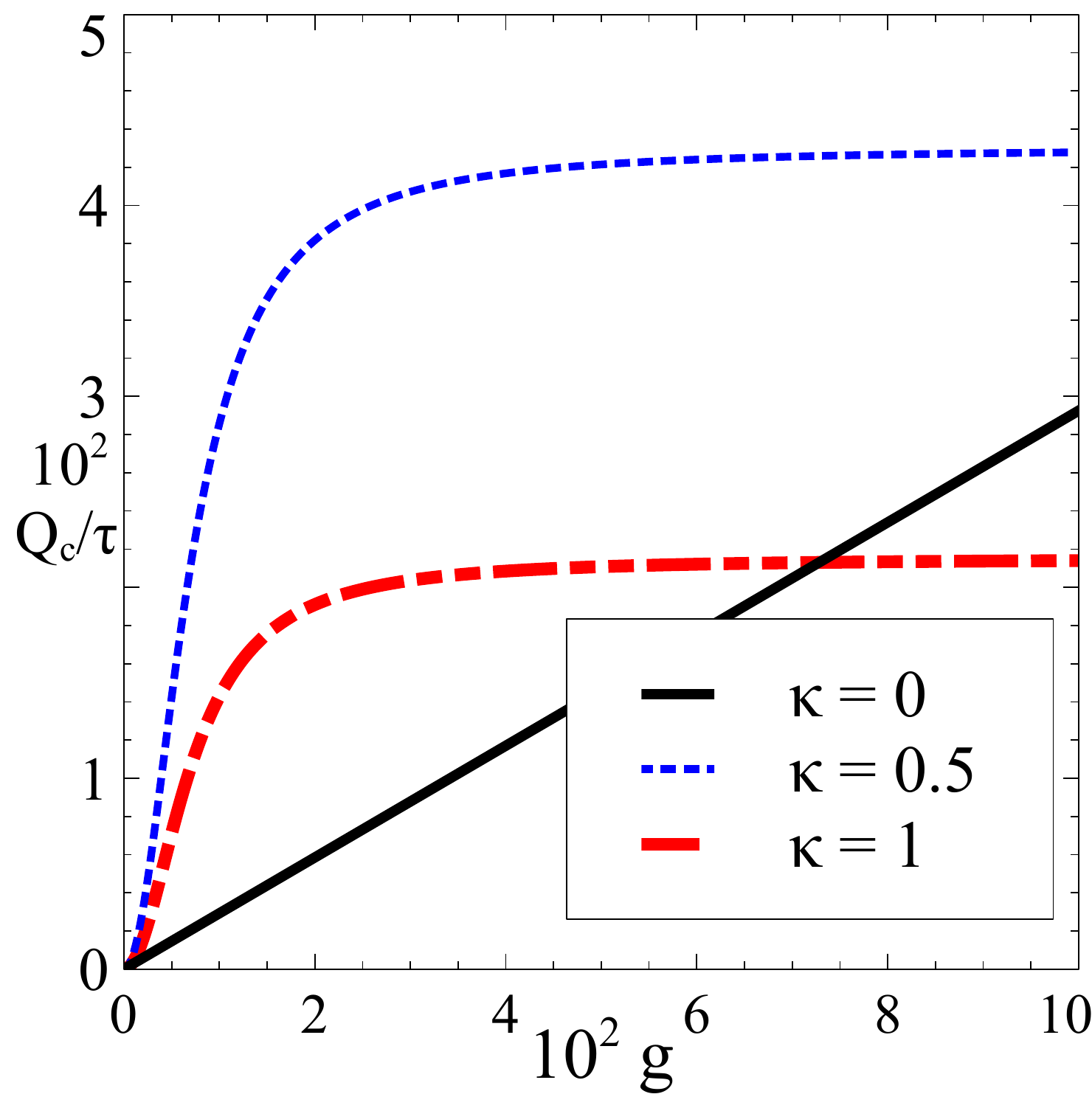}       
    \label{with_coh_g_1} \quad
  \includegraphics[width=0.23\textwidth, keepaspectratio]{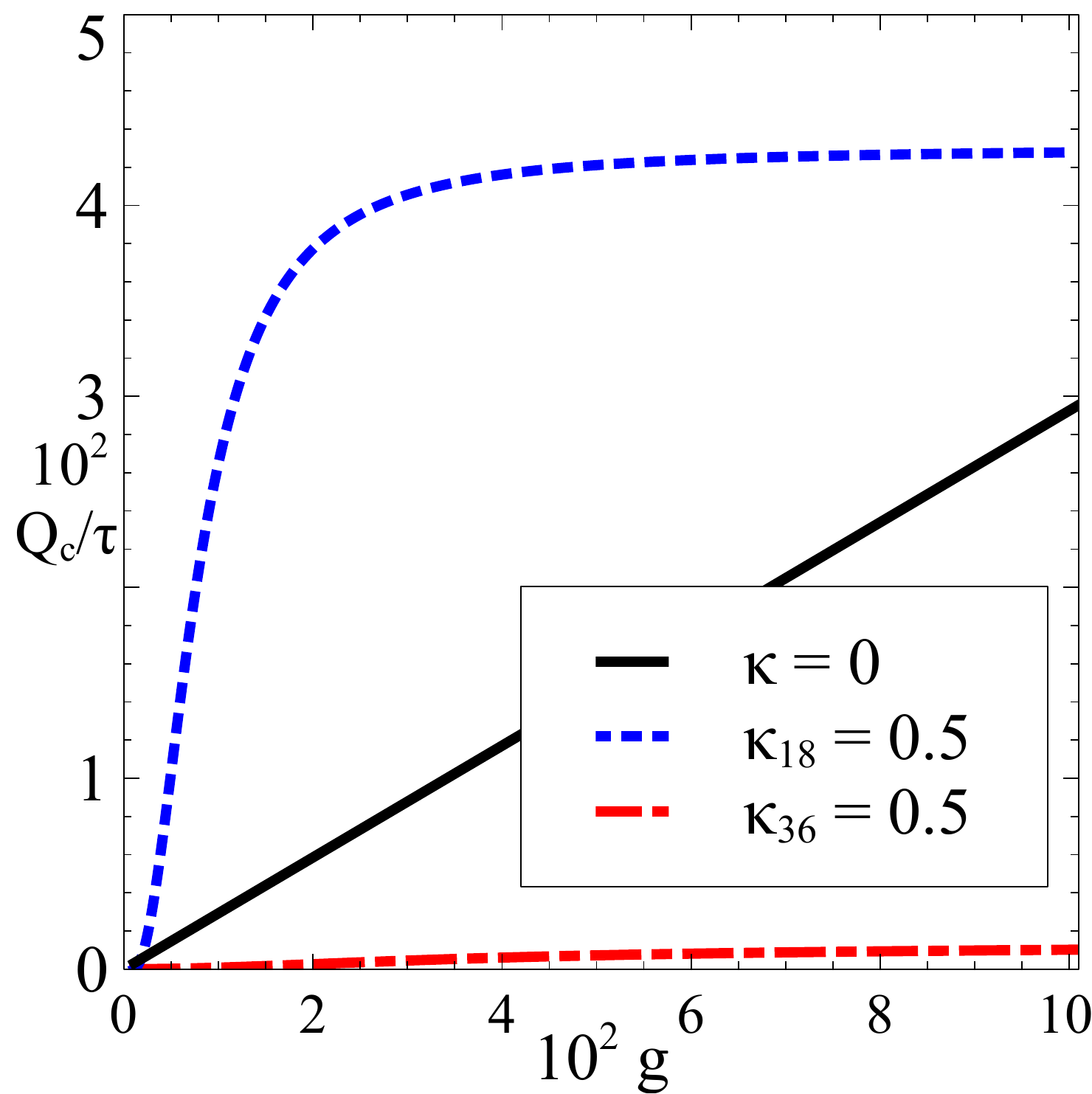}       
    \label{with_coh_g_2}
\caption{ Variation of BSOCR with interaction strength $g$. \textit{Left:} for different strengths of initial coherence $\kappa$ applied to the $|000\rangle\langle111|$ subspace as well as  \textit{Right : }for same amount of coherence in different subspaces viz. $36$, i.e., $|101\ket\bra 010|$, and $18$, i.e., $|000\ket\bra 111|$ subspaces. The reset probabilities have been taken as $p_c = p_r = p_h = p$ throughout. The other parameters are taken as $T_c = 1, T_r = 2, T_h = 10 , \eta = 0.5 ,$ $ p = 0.05, E_c = 1$.}
\label{g_with_coherence}
\end{figure}

In Fig. \ref{g_with_coherence}, we demonstrate the linear dependence of the BSOCR on the interaction strength. It will be interesting to inquire whether some initial coherence in the three qubit QAR can boost the BSOCR beyond the linear increase with the interaction strength $g$ seen above. We add an additional real off diagonal part to the $|000\ket\bra 
111|$ and the corresponding adjoint element  of the diagonal initial 
density matrix $\rho_{0}$ with the following magnitude \beq = \kappa 
\sqrt{\prod_{i=c,r,h} r_{i} \bar{r_{i}}}\quad ; \quad 0 \leq \kappa \leq 1.  \label{inputcoh}\eeq Fig.  \ref{g_with_coherence} shows that
this initial coherence
can significantly increase the BSOCR. Thus, quantum coherence, already identified as a useful resource in quantum
information theory and quantum thermodynamics \citep{adessoreview, spekkens, avijit, lostaglio}, can enhance the performance of the QAR for a fixed interaction strength by reducing the minimum time taken to reach the steady state.
It is worth mentioning that in Ref. \citep{Mitchison}, it has been pointed out that coherence can enhance the cooling of
the cold qubit in the 
transient regime. In this work, we show that coherence can also enhance the performance of the steady heat machine. This is consistent with the assertion made in Ref. \cite{Debasis} that quantum coherence can serve to augment the speed limit for general dynamics. {\color{black} At this juncture, we want to emphasize that the present treatment of adding initial coherence to the system is different from having a bath which is coherent. The choice of adding coherence only to the $|000\ket\bra 111|$ subspace may seem restrictive, but it can be shown that the nature of functional dependence of the BSOCR on the interaction strength or reset probabilities does not change whether we add the coherence in any other density matrix element, say, $|010\ket\bra101|$.  However, the numerical value of the BSCOR depends on the subspace to which coherence is added. In Fig. \ref{g_with_coherence}, we show that as far as the efficacy of applying coherence to facilitate cooling is concerned, applying coherence to $|000\ket\bra111| $ subspace is far better than applying the same amount of coherence to $|010\ket\bra101|$ subspace.}

%

%

\subsection{Thermalization strength vs BSOCR}
\begin{figure}
\centering
\includegraphics[width=0.23\textwidth, keepaspectratio]{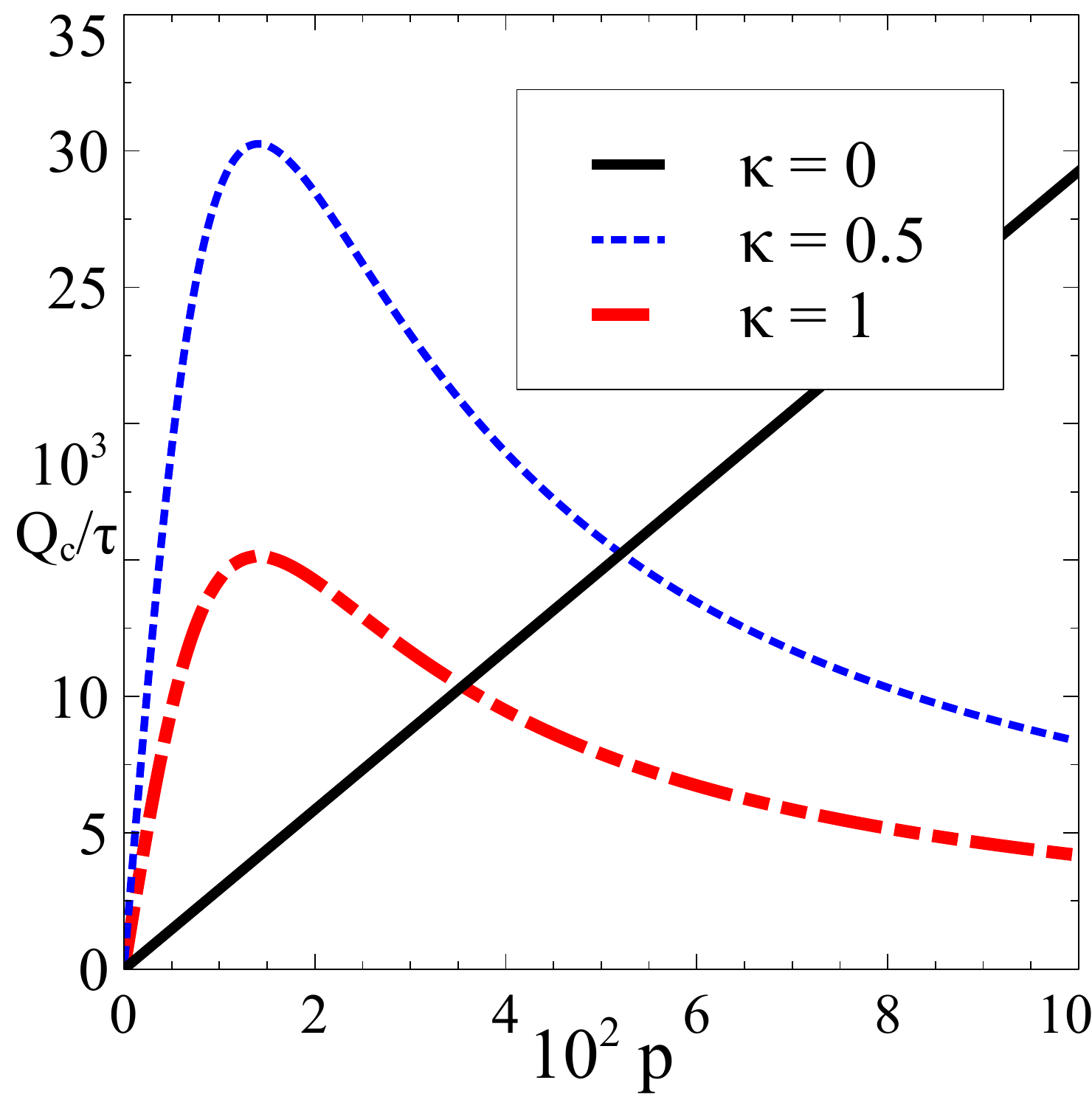}       
    \label{with_coh_p_1} \quad
    \includegraphics[width=0.23\textwidth, keepaspectratio]{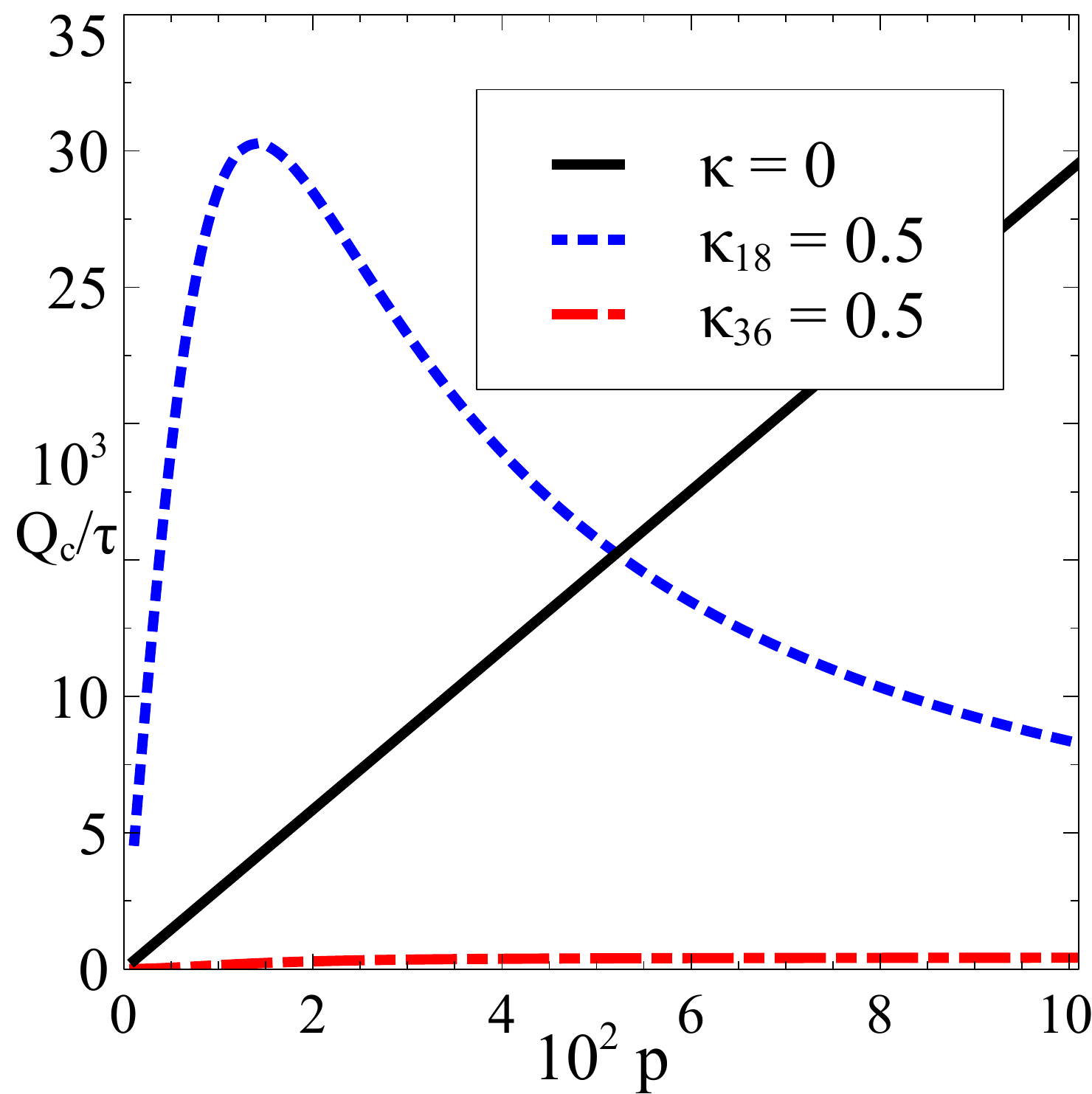}       
    \label{with_coh_p_2}
\caption{ Variation of BSOCR with reset probability $p_{c} = p_{r} $
= $p_{h}$ $= p$. \emph{Left :} for different strengths of initial coherence $\kappa$ in the $|000\rangle\langle 111|$ subspace as well as \emph{Right :} for same amount of coherence in different subspaces viz. $36$, i.e., $|101\ket\bra 010|$, and $18$, i.e., $|000\ket\bra 111|$ subspaces. The other parameters are taken as $T_c = 1, T_r = 2, T_h = 10, \eta = 0.5, g = 0.05, E_c = 1$.}
\label{zeno_with_coherence}
\end{figure}

In this section we study the dependence of the BSOCR on the thermalization strengths, i.e., the reset probabilities.
In case the reset probabilities are different, the expression for BSOCR is a bit 
cumbersome. Thus we confine ourselves to the case of equal reset probabilities. If the reset probabilities vanish, there is no energy exchange and consequently no cooling either. \emph{If the reset probabilities $p_{c},p_{r}, p_{h}$ are all equal to, say $p$, then 
the BSOCR is linearly proportional to $p$.} Again, one can verify that this simple linear relation does not hold for the 
heat extracted from the cold bath. Now, let us start with a slightly more general case  where the density matrix corresponding to the initial state $\rho_{0}$ possesses some non-local  coherence, the latter typified by non-zero elements as in Eq. \ref{inputcoh} in the earlier section. It is readily verified that the reduced states are still Gibbsian in their respective energy eigenbases. Fig. \ref{zeno_with_coherence} shows that 
although the BSOCR increases rapidly with increasing reset probability if the 
$p$ is very low, it decays considerably as we go on increasing $p$. Of course we cannot indefinitely go on increasing the reset probability due to the weak coupling assumption used in the derivation of this master equation but the decrease of the BSOCR with increasing reset probability is observed even in the regime depicted in the figure, where $g, p \ll E_{1} $. 

\subsection{Efficiency vs BSOCR}
\begin{figure}
\includegraphics[scale=0.4]{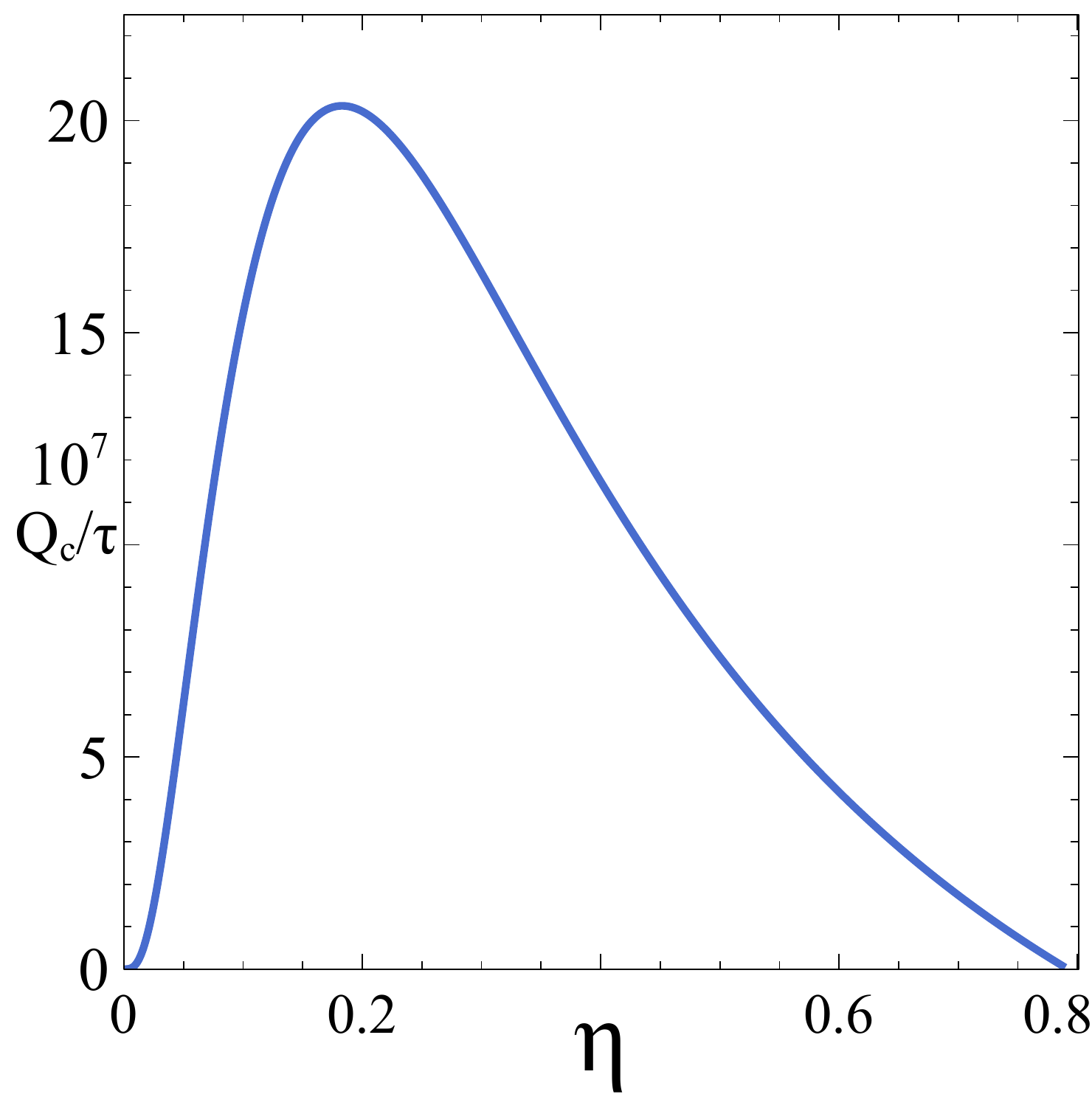}
\caption{Variation of BSOCR with steady state efficiency $\eta$ 
for fixed parameter values $p_c = 0.01, p_r = 0.02, p_h = 0.05, T_c = 1, T_r = 2, T_h = 10, E_c = 1, g = 0.01$.}
\label{pow_vs_eta}
\end{figure}
For finite time thermodynamic cycles, finding the efficiency at maximum power is a well-motivated pursuit. For QAR models, in addition to finding the efficiency at maximum power \citep{adesso1}, it makes sense to find the efficiency of the QAR when the BSOCR, the figure of merit encompassing both steady power characteristics and speed of evolution, is maximal.  Let us now focus on how the BSOCR depends on the efficiency of the machine. For this model, the efficiency $\eta$ is expressed as $E_{c}/E_{h}$ .  This allows us to explicitly compute the BSOCR for any 
efficiency. We demonstrate in Fig. \ref{pow_vs_eta} that in a generic case, the BSOCR vanishes when there is no cooling as well as at the Carnot point, and attains its maximal value at some intermediate point.  Exactly calculating the efficiency at maximal BSOCR in the most general case is quite cumbersome, therefore we restrict ourselves to the case of equal reset probabilities and use the expression for BSOCR derived in the supplementary material \citep{supple}.  In the high temperature limit and with the following assumptions, this allows us to derive the following crucial result, the proof of which can be found in the supplementary material \citep{supple}. 

\emph{\textbf{Theorem} - If the 
reset probabilities $p_{c},p_{r}, p_{h}$ are equal and $T_{h} \gg T_{r} \gg
T_{c}  \gg E_{c}$  along with the condition that $\frac{T_{c}}{T_{r}} \approx 
\frac{T_{r}}{T_{h}}$, the efficiency at maximal BSOCR equals 
\beq \frac{T_{c}}{T_{r}}\left(1-\sqrt{\frac{T_{c}}{T_{h}}}\right).
\label{etamax}
\eeq}

\noindent Let us mention that in the limit considered above, the Carnot efficiency for this refrigerator is given by

\beq
\eta_{Carnot} = \frac{T_{c}}{T_{r}} \left( 1 - \frac{T_{c}}{T_{h}} \right).
\label{etacarnot}
\eeq
We immediately see that the relation between Eq. \ref{etamax} and Eq. \ref{etacarnot} is remarkably similar to the relation between the expressions for Carnot bound and  Curzon-Ahlborn (CA) bound for efficiency heretofore derived for various cyclic heat pumps \citep{abah, yan}. 

\section{conclusion}
We have introduced a figure of merit, which we call the BSOCR, defined as the product of the steady cooling power and the maximum speed of evolution towards the steady state, and analyzed it for a specific model of QAR. We have shown that the BSOCR scales linearly with the interaction strength.  We have also observed that the BSOCR increases  with the value of reset probabilities. Consequently, we have observed how the relation of the BSOCR with interaction strength and reset probabilities change with the introduction of initial quantumness in the system. Interestingly, we have demonstrated that the initial quantumness is helpful to increase the magnitude of BSOCR. Strikingly, we have shown that the efficiency at maximum BSOCR  has a simple expression reminiscent of the CA type expressions in a limiting case. That the steady state performance parameter depends on the dynamics which generates the steady state is perhaps worth exploring  further for more generic heat machines.  Further work elaborating on the link between cycle based and self contained thermal machines may elucidate this connection even more. In place of the speed limit used in this work, one may put alternate expressions for the speed limit to derive alternate forms of the BSOCR. Further work on power characteristics of general quantum thermal devices as well as quantum biological processes using the procedure outlined here may be performed. {\color{black} Another potentially interesting avenue of future investigation is to consider non-Markovian baths, e.g., Fermionic central spin baths, and see whether they impart any advantage as far as the BSOCR is concerned. The seminal paper by Deffner and Lutz \citep{deffnerlutz} indeed demonstrates that the time taken to reach a target state may be significantly less for non-Markovian processes. Thus, if a steady state exists for a non-Markovian model, the value of our figure of merit may be significantly enhanced. One particular obstacle towards pursuing this idea in a straighforward way is the fact that for such baths - there may not even exist any unique steady state in general. We are optimistic that engaging non-equilibrium steady states in the discussion of quantum autonomous refrigators may be a very fruitful approach (see, for example, Ref. \cite{ness} for such a recent work). The treatment of such situations lies outside the scope of the present paper, but may be considered as important future projects. } At the time of completion of the first draft of the present manuscript, two recent preprints \citep{csl1,csl2} appeared which proposed the existence of speed limits to evolution in the classical phase space. It may be interesting to perform a similar analysis to ours for classical heat machines. 


The authors acknowledge financial support from Department of Atomic Energy, 
Govt. of India.

\section*{Appendix}

In the appendix, we present the calculations for the results in the main work. The initial state of the three-qubit refrigerator is the product of locally thermal states $ \rho_{0} = \tau_{c} \otimes \tau_{r} \otimes \tau_{h} $ and the steady state is given by $\rho_{f} = \rho_{0} + \gamma \sigma_{crh} $, as described in the main paper.  First, let us note that the cooling rate $Q_c$ is expressible in the form \beq Q_c = \frac{\xi_1}{\Upsilon_1 + \frac{\Upsilon_2}{g^2} }  \ \eeq and the inverse speed limit, i.e. minimum time required for evolution to the steady state is similarly expressible in the form \beq \tau  = \frac{\xi_2 /g }{\Upsilon_1 + \frac{\Upsilon_2}{g^2} } , \ \eeq where the parameters $\Upsilon_1, \Upsilon_2, \xi_1, \xi_2$ all depend on the system parameters other than g. All these parameters are explicitly expressed as $\xi_1 = q E_c \Delta, \xi_2 = \frac{\Delta \tr (\tau_c \tau_r \tau_h \sigma_{crh})}{\sqrt{2} (r_1 r_3 + r_2 (r_1 + r_3 -2 r_1 r_3 -1))}, \Upsilon_1 = 2 +\sum_i q_i + \sum_{jk} Q_{jk} \Omega_{jk}, \Upsilon_2 = q^2 /2 $. Thus, if other system parameters are kept fixed and only the interaction strength $g$ is tuned, then the above parametric relation yields the following link between $Q_c$ and $\tau$, which is valid for any interaction strength. 

\beq \tau^2 = \left(\frac{\xi_2 ^2}{\xi_1} \right) Q_c  - \Upsilon_1 \frac{\xi_2^2}{\xi_1^2} Q_c^2\eeq

In case the first term of the RHS dominates over the second term, this immediately reveals that there exists a trade-off between the speed of evolution and the steady state cooling rate, which provides the rationale for investigating this tradeoff through our figure of merit. In this sense, the existence of this tradeoff is not obvious from the outset. However,  in general, if the cooling rate $Q_c$is small, the quadratic term is subleading, and the tradeoff is observed. More specifically, in the scenarios we illustrated in the paper, for example in Fig. 1 of the main text, the first term indeed dominates and this trade-off is observed.

The BSOCR $\chi$  is the ratio of the cooling power at steady state and the minimal time of evolution from $\rho_{0}$ to $\rho_{f}$, i.e.,
\begin{equation}
 \chi =  q \gamma E_{c}  \frac{\sqrt{ \tr  
\left(\mathcal{L}^{\dagger} \rho_{0} \right)^{2}}}  {|\cos \theta -1| \tr \rho_{0}^{2}}
\label{bsocr_appendix}
\end{equation}
The relative purity angle $\theta$ between initial and final states is given by $\theta = \cos^{-1} \frac{\tr \left(\rho_{0} \rho_{f} \right)}{ \tr 
\left(\rho_{0}^{2} \right)}$. Consequently, $|\cos \theta - 1| \tr (\rho_{0}^{2})$ equals $|\gamma \tr (\rho_{0} \sigma_{crh})| $. The conjugate of Lindbladian acting on the initial state is simplified as 
\bea \mathcal{L}^\dagger \rho_{0} =  -i [H_{int}, \rho_{0}] + \sum_{i} p_{i} \left( \tau_{i} \otimes \tr_{i} \rho_{0} - \rho_{0}  \right).
=  -i [H_{int}, \rho_{0}]  \eea
\\
The expression for BSOCR thus equals $\frac{q E_{c} 
\sqrt{\tr [H_{int} , \tau_{c} \tau_{r} \tau_{h}]^{2}}}{\tr \left(\tau_{c} 
\tau_{r} \tau_{h} \sigma_{crh} \right) } $.
For equal reset probabilities, i.e., on putting $p_{c} = p_{r} = p_{h} = 
p$, the expression for BSOCR written in terms of the excited state probabilities $\lbrace \bar{r}_i \rbrace$ reduces to
\begin{widetext}
 \footnotesize 
\bea \frac{Q_{c}}{\tau} = 3\sqrt{2}g p 
E_{c}\frac{\bar{r}_{c} r_{h} - \bar{r}_{r} \left(1-\bar{r}_{c} - \bar{r_{h}} + 2 \bar{r}_{c} \bar{r}_{h} 
\right)}{\frac{3}{2} - 3 \bar{r}_{h} + \bar{r}_{r} (-1 + 3 \bar{r}_{h} + \bar{r}_{h}^{2} ) + \bar{r}_{r}^{2} (5 
- 9 \bar{r}_{h} + 4 \bar{r}_{h}^{2}) + \bar{r}_{c}^{2} (\bar{r}_{r} + \bar{r}_{r}^{2} (4-8 \bar{r}_{h}) + 8 \bar{r}_{r} 
\bar{r}_{h}^{2} - \bar{r}_{h} (1+ 4 \bar{r}_{h})) - \bar{r}_{c} (3 - 5 \bar{r}_{h} + \bar{r}_{h}^{2} + 6 \bar{r}_{r} \bar{r}_{h} 
- 3 \bar{r}_{r} + \bar{r}_{r}^{2} (9-16 \bar{r}_{h} + 8 \bar{r}_{h}^{2})) }. \nonumber \\ \label{damra} \eea 
\end{widetext}
\normalsize
\subsection{BSOCR in presence of initial coherence}
We add an additional real positive off diagonal part the $|000\ket\bra 
111|$ and the corresponding adjoint element  of the diagonal initial 
density matrix $\rho_{0}$  with the following magnitude \[ = \kappa 
\sqrt{\prod_{i=c,r,h} r_{i} \bar{r_{i}}}\quad ; \quad 0 \leq \kappa \leq 1. \]

Now, the corresponding BSOCR is given by
\beq
\frac{Q_{c}}{\tau} = \frac{qE_{c} \sqrt{\tr \left(-i [H,\rho_{0} + \mu] - q \mu 
\right)^{2}}}{\left|\tr\left(\rho_{0} \sigma + \mu \sigma \right) - 
\frac{1}{\gamma} \tr \left(\rho_{0}\mu + \mu^{2} \right) \right|},
\eeq
where $\mu =\kappa \sqrt{\prod_{i=c,r,h} r_{i} \bar{r_{i}}} \left( 
|0_{c}0_{r}0_{h}\ket\bra 1_{c}1_{h}1_{r}| + h.c. \right) $. Explicitly computing this expression yields the following expression for BSOCR 
\beq = 3p E_{c} \sqrt{\frac{N_{1} p^{2} + N_{2} p + N_{3}}{D_{1} p^{4} + D_{2} 
p^{2} + D_{3}}}, \eeq 
where 
\begin{widetext}
 \beq N_{1} = 9\tr \mu^{2}, N_{2} = -6 \tr M\mu, N_{3} = \tr M^{2}, D_{1} 
= 81 \frac{\Pi_{2}^{2}}{4 g^{4} \Delta^{2}} , D_{2} = 18 \left(\Pi_{1} + 
\frac{\lambda \Pi_{2}}{\Delta} \right) \frac{\Pi_{2}}{2 g^{2} \Delta}, D_{3} = 
\left(\Pi_{1} + \frac{\lambda \Pi_{2}}{\Delta} \right)^{2},  \lambda = 2+ \sum_i q_i + \sum_{jk} Q_{jk}\Omega_{jk}. \eeq 
\end{widetext}
Here, \beq 
M = -i [H,\rho_{0} + \mu], \gamma = \frac{-\Delta}{\lambda + 
\frac{9p^{2}}{2g^{2}}}, \Pi_{1} = \tr \left( \rho_{0} \sigma + \mu\sigma 
\right), \Pi_{2} = \tr (\rho_{0} \mu + \mu^{2}). \eeq

\noindent Clearly only $N_{3}$ and $D_{3}$ coefficients survive in the absence of initial coherence, i.e., $\kappa = 0$, thus giving rise to the linearity with $p$ in that case. One may easily check now for $\mu = 0$ that $\frac{N_{3}}{D_{3}}$ is proportional to $g^{2}$, thus confirming the linearity of BSOCR with interaction strength. In fact it may be shown that the BSOCR can be expressed in the following 
alternate way \beq  = g E_{c}   \sqrt{\frac{N'_{1} g^{2} + N'_{2} g + 
N'_{3}}{D'_{1} g^{4} + D'_{2} g^{2} + D'_{3}}}, \eeq with only the coefficients 
$N'_{3}, D'_{3}$ surviving in the special case of $\kappa = 0$, i.e.,  no initial 
coherence.

\subsection{Derivation of efficiency at maximum BSOCR}

\emph{\textbf{Efficiency at maximal BSOCR in the high temperature limit-} If the 
reset probabilities $p_{c},p_{r}, p_{h}$ are equal and $T_{h} \gg T_{r} \gg
T_{c}  \gg E_{c}$  along with the condition that $\frac{T_{c}}{T_{r}} \approx 
\frac{T_{r}}{T_{h}} $ , the efficiency at maximal BSOCR equals 
$\frac{T_{c}}{T_{r}}\left(1-\sqrt{\frac{T_{c}}{T_{h}}} \right)$. }

\textit{\textbf{ Proof :}}  
In the high temperature limit, the occupancies of the excited states of qubits are approximated by  \beq \bar{r}_{i} \cong \frac{1}{2} - 
\frac{x_{i}}{4} - \frac{x_{i}^{2}}{8} + \frac{x_{i}^{3}}{4} 
\hspace{0.25in};\hspace{0.2in} x_{i}  = \frac{E_{i}}{T_{i}}. \eeq   Putting them 
in the expression for BSOCR and weeding out higher order terms using the fact 
that   $T_{h} \gg T_{r} \gg T_{c}  \gg E_{c}$ leads to the following 
reasonably simple expression for the BSOCR 
\beq \frac{Q_{c}}{\tau} \cong \frac{3\sqrt{2}g p E_{c}}{3 - \frac{x_{c} x_{r} 
(x_{c} - x_{r})}{3(x_{c} - x_{r} + x_{h}) }} = \frac{3\sqrt{2}g p E_{c}}{3 - 
\frac{1}{3}F}.
\eeq

\noindent Now optimizing the BSOCR reduces to maximizing $F$. Putting $x_{c} = 
\frac{E_{c}}{T_{c}} , x_{r} = \frac{E_{c}}{T_{r}} \left(1+\frac{1}{\eta} \right) 
, x_{h} = \frac{E_{c}}{\eta T_{h}}$ , we differentiate $F$ with respect to 
$\eta$ to obtain the efficiency at maximal BSOCR. 

The solution to the equation $\frac{\partial F }{\partial \eta} = 0$ yields the 
following nontrivial solutions for efficiency at extremal BSOCR - 

\beq \eta_{opt} =  \frac{-T_{c}^{2}T_{r} + T_{c}^{2} T_{h} - T_{c} T_{r} T_{h} 
\pm \sqrt{T_{c}^{4} T_{r}^{2} + T_{c}^{3} T_{r}^{2} T_{h}}}{2 T_{c}^{2} T_{r} - 
T_{c}T_{r}^{2} - T_{c}^{2} T_{h} + 2 T_{c}T_{r} T_{h} - T_{r}^{2} T_{h} }. \eeq 

\noindent Applying the condition $T_{h} \gg T_{r} \gg T_{c}$ , we now have the 
expression for efficiency at extremal BSOCR as 

\beq 
\eta_{opt} \cong \frac{T_{c} T_{r} T_{h} - T_{c}^{2} T_{h} \mp \sqrt{T_{c}^{3} 
T_{r}^{2} T_{h}}}{(T_{r} - T_{c})^{2} T_{h}} = \frac{T_{c}}{T_{r} - T_{c}} \mp 
\frac{T_{c}T_{r}}{(T_{r} - T_{c})^{2}} \sqrt{\frac{T_{c}}{T_{h}}},
\eeq 

\noindent Noting $T_{r} - T_{c} \approx T_{r} $, we arrive at the following expression for 
efficiency at optimal BSOCR 

\beq \eta_{opt} = \frac{T_{c}}{T_{r}} \left( 1 \mp \sqrt{\frac{T_{c}}{T_{h}}} 
\right).  \eeq

\noindent Now one can show that the minus sign corresponds to the \emph{maximal} BSOCR 
(taking into account the fact  $\frac{T_{c}}{T_{r}} \sim \frac{T_{r}}{T_{h}} $, the other solution lies beyond the Carnot efficiency) and consequently the 
efficiency at maximal BSOCR is given by 
$\frac{T_{c}}{T_{r}}\left(1-\sqrt{\frac{T_{c}}{T_{h}}} \right)$ - thus completing 
the proof. \qed \\
One can check that the Carnot efficiency $\eta_{Carnot}$ is given by \[ 
\eta_{Carnot} = \frac{T_{c}}{T_{r}} \left[ 1 - \frac{T_{c}}{T_{h}} + \left( 
\frac{T_{c}}{T_{r}} - \frac{T_{r}}{T_{h}}  \right)\right]. \]  If 
$\frac{T_{c}}{T_{r}} \approx \frac{T_{r}}{T_{h}} $ , the Carnot efficiency is 
given by 
 
\[ \eta_{Carnot} = \frac{T_{c}}{T_{r}} \left(1- \frac{T_{c}}{T_{h}} \right). \]

\noindent One can compare this with the efficiency at maximal BSOCR obtained here , which 
is given by -

\[ \eta_{opt} =  \frac{T_{c}}{T_{r}} \left(1- \sqrt{\frac{T_{c}}{T_{h}}} \right) .
 \]

\bibliographystyle{apsrev4-1}
\bibliography{fridge_1}

\end{document}